\documentclass[%
  twoside,
  reprint,
  amsmath,amssymb,
  aps,
  pra,
  nofootinbib,
  a4paper
]{revtex4-1}

\usepackage{graphicx}
\usepackage{dcolumn}
\usepackage{bm}
\usepackage[usenames,dvipsnames]{xcolor}
\usepackage{textcomp,gensymb} 
\usepackage{subfigure}

\usepackage{tikz}
\usetikzlibrary{arrows}
\usepackage{MnSymbol} 
\usepackage{rotating}
\usepackage{relsize}

\usepackage[
text={7.3in,10in},centering,
total={6.5in,8.75in}, top=1in, left=0.52in, includefoot,
]{geometry}

\usepackage[
  bookmarks=true,
  colorlinks,
  linkcolor=blue,
  urlcolor=blue,
  citecolor=blue,
  plainpages=false,
  pdfpagelabels,
  final,
  breaklinks=true
]{hyperref}
\hypersetup{
pdftitle={Spin conservation in high-order-harmonic generation using bicircular fields}, 
pdfauthor={Emilio Pisanty, Suren Sukiasyan and Misha Ivanov}
}

\usepackage{natbib}
\makeatletter \def\NAT@def@citea{\def\@citea{\NAT@separator\,}} \makeatother

\newcommand{\vb}[1]{\mathbf{#1}}
   \newcommand{\vbr}{\vb{r}}
   \newcommand{\vbk}{\vb{k}}
   \newcommand{\vbe}{\vb{E}}
   \newcommand{\vbp}{\vb{P}}
\newcommand{\une}{\hat{\vb{e}}}

\newcommand{\cc}{\text{c.c.}}
\newcommand{\eps}{\varepsilon}

\newcommand{\chit}{\stackrel{\leftrightarrow}{\chi}\!\!{}^{(3)}}
\newcommand{\chis}{\chi^{(3)}_\text{s}}

\usepackage{stackengine} 
\newcommand{\tdots}{\setstackgap{S}{0.25ex}\!\!\mathrel{\Shortstack{{.} {.} {.}}}\!}

\newcommand{\leftpol}{\!\; \!\!\rcirclearrowright}
\newcommand{\rightpol}{\!\; \!\!\lcirclearrowright}
\newcommand{\epol}{\resizebox{0.8em}{1em}{$\lcirclearrowright$}}

\newcommand{\etal}{\textit{et~al.}}

\begin{document}

\title{Spin conservation in high-order-harmonic generation using bicircular fields}

\author{Emilio Pisanty$^{1}$}
 \email{e.pisanty11@imperial.ac.uk}
\author{Suren Sukiasyan$^{1}$}
\author{Misha Ivanov$^{1,2,3}$}%
 \email{m.ivanov@imperial.ac.uk}
\affiliation{%
\scriptsize
$^1$Blackett Laboratory, Imperial College London, South Kensington Campus, SW7 2AZ London, United Kingdom\\
$^2$Department of Physics, Humboldt University, Newtonstrasse 15, 12489 Berlin, Germany\\
$^3$Max Born Institute, Max Born Strasse 2a, 12489 Berlin, Germany
}

\date{\today}

\begin{abstract}
We present an alternative theoretical model for a recent experiment [A.~Fleischer~\etal, \href{http://dx.doi.org/10.1038/nphoton.2014.108}{\textit{Nature Photon.} \textbf{8}, 543 (2014)}] which used bichromatic, counter-rotating high intensity laser pulses to probe the conservation of spin angular momentum in high harmonic generation. We separate elliptical polarizations into independent circular fields with definite angular momentum, instead of using the expectation value of spin for each photon in the conservation equation, and we find good agreement with the experimental results. In our description the generation of each individual harmonic conserves spin angular momentum, in contrast to the model proposed by Fleischer \etal{} Our model also correctly describes analogous processes in standard perturbative optics.
\end{abstract}

\maketitle

\section{Introduction}

The process of high harmonic generation~\cite{HHGTutorial,*JoachainReview} is the flagship experiment of extreme nonlinear optics. It consists of irradiating atoms or small molecules with long-wavelength laser pulses whose electric field is comparable to the internal electric fields of the atoms, which results in the emission of harmonics of the driving laser field that can span several thousands of orders~\cite{UltrahighHG}, most frequently with a flat plateau in their intensity. It can be understood intuitively in terms of a three-step model where an electron is tunnel ionized, propagates classically in the laser field away from the ion and back, and recombines with the ion upon recollision, emitting a burst of high-frequency radiation~\cite{HHGTutorial,*JoachainReview}.

In the final recombination step, the electron rejoins the ion by re-filling the atom it originally left behind, which leaves the atom in its ground state. Thus, although it is usually accompanied by ionization and other processes, high harmonic generation (HHG) is typically seen as a parametric process in which the initial and final states are the same. As a parametric process, it must obey conservation laws for energy, momentum, and orbital angular momentum, and these have been successfully demonstrated in the laboratory \cite{EnergyConservationExperiment, MomentumConservationExperiment, OAMConservationExperiment}.

A recent, ingeniously conceived experiment \cite{FleischerCohen}, probes whether the process conserves spin angular momentum. In this experiment, argon atoms are subjected to a superposition of two counter-rotating laser fields of different frequencies~\cite{EichmannExperiment, SFALong, SFAMilosevic, SFAMilosevicBecker, VitaliSelectionRules, SFACeccherini, MilosevicIsolatedPulses}. This permits the use of incident light with spin whilst avoiding the dipole selection rules that forbid harmonic emission in a single-color circularly-polarized field. In this setup harmonics are produced with nearly the same intensity as for linear fields, with the additional control of the harmonics'  polarization through changes in the ellipticity of the driving field.

Fleischer \etal~\cite{FleischerCohen} have provided a simple model based on perturbative optics which explains the essentials of the spectra they observe, and which shows that the harmonic generation process, as a whole, does indeed in many cases conserve the spin angular momentum of light. However, they argue that some situations require the electron to carry away angular momentum after the recollision, and that in general the production of each individual harmonic is not a closed process.

This model, which we call Model~1, suggests that spin angular momentum is not conserved for certain harmonic lines, so that some harmonics must be assumed to be emitted in correlated pairs for the overall process to be parametric. Additionally, a `strong field correction' is introduced empirically, and for certain harmonics the model's predictions depend discontinuously on the experimental parameters in a way we find unphysical, and which is not present in the description of lower-ordered processes which are well understood.

In this paper we present an alternative perturbative-optics model for this experiment. In essence, we posit that, as regards nonlinear optics, elliptically polarized fields should be seen as the superposition of circularly polarized fields of different amplitudes which contribute photons of definite spin $\sigma=\pm1$, instead of single photons that contribute their expectation value for the spin, $|\langle\sigma\rangle|<1$. This model accounts for the same experimental results as Model~1 without any free parameters, and it does not have unphysical discontinuities. Further, it provides specific predictions that can be tested numerically (and, in principle, experimentally), by using rotating elliptical polarizations whose left- and right-circular components are slightly detuned. Within this model, the generation of each harmonic is a closed process that does conserve spin angular momentum in all cases.

This paper is structured as follows. In \ref{sec:review} we review the essentials of the experiment and of Model~1. In \ref{sec:model} we pre\-sent our own model, Model~2, and explain its differences to Model~1. In \ref{sec:SubchannelSplittings} we explore the predictions of Model~2 for rotating elliptical polarizations, and in \ref{sec:perturbativesetting} we apply both models to the lowest-order process, sum-frequency generation in four-wave-mixing, which embodies and highlights the differences between them.

\section{The experiment}
\label{sec:review}
The experiment of Fleischer \etal~\cite{FleischerCohen} uses two co-pro\-pa\-ga\-ting laser drivers of equal intensity. One is centred at 800 nm and the other, at 410 nm, is obtained from the longer wavelength by a red-shifted second harmonic generation setup; the slightly-off-integer ratio between the frequencies is used to identify how many net photons from each field have been absorbed. Both linearly-polarized drivers go through quarter-wave plates which are free to rotate independently, and the drivers are then combined. 

The resulting electric field performs a variety of Lissajous figures in the polarization plane, which slowly drift throughout the pulse due to the slight detuning between the drivers. The pulses comprise about fourteen cycles of the fundamental and thus equally many passes of the Lissajous figure. In the `bicircular' setting, with both pulses fully circularly polarized, the Lissajous figure is a trifolium. The harmonic spectrum in this case covers the integer orders not divisible by three.

The experimental observations consist of scans over the ellipticity of each of the drivers while the other is held constant at the circular polarization. This opens up a variety of harmonic channels, including channels at orders divisible by 3, which are otherwise forbidden by symmetry. These subsidiary channels are slightly detuned from the main ones, which is due to the slightly off-integer ratio $r=1.95$ between the frequencies of the two drivers. This detuning enables a unique assignment of integers $n_1$ and $n_2$ of photons absorbed from each driver, in terms of which the frequency of each channel is
\begin{equation}
 \Omega_{(n_1,n_2)}=n_1\omega+n_2 r\omega,
\end{equation}
where $n_1+n_2$ must be odd by conservation of parity.

In Model~1, each driver photon is considered to contribute to the process angular momentum equal to the field's expectation value of spin angular momentum $\langle\sigma_j\rangle$. One then expects this total angular momentum, $n_1\langle\sigma_1\rangle+n_2\langle\sigma_2\rangle$, to match the angular momentum of the outgoing harmonic photon.

In the bicircular setting, the fundamental driver's polarization is right circular, with complex unit vector $\une_{R}=\tfrac{1}{\sqrt{2}}(\une_H+i\une_V)$, and its photons have definite spin $\sigma_1=+1$. Similarly, the harmonic driver is left-circular polarized along $\une_L$, and its photons have definite spin $\sigma_2=-1$. In the general case each driver is elliptically polarized and can be written in the form 
\begin{equation}
 \vbe=
 \frac{E_0e^{-i\omega t}}{2\sqrt{2}}
 \left(
 \frac{1+\eps}{\sqrt{1+\eps^2}}\une_R
 +
  \frac{1-\eps}{\sqrt{1+\eps^2}}\une_L
 \right)
 +\cc,
 \label{EllipticalAsSumOfCircularPolarizations}
\end{equation}
where $\eps\in[-1,1]$ is the signed ellipticity of the field. The expected angular momentum of this field can be calculated to be
\begin{equation}
 \langle\hat{\sigma}\rangle=\frac{2\eps}{1+\eps^2}
 \label{ExpectedAngularMomentumEllipticity}
\end{equation}
in units of $\hbar$. For a field generated by shining linearly po\-la\-rized light on a half-wave plate at an angle $\alpha$ to its fast axis, as in the experiment \citep{FleischerCohen}, the ellipticity thus reduces to 
\begin{equation}
 \langle\hat{\sigma}\rangle=\sin(2\alpha).
 \label{ExpectedAngularMomentumFromAngle}
\end{equation}

Under these assumptions, the conservation equation can now be formulated: the spin of the resulting harmonic photon on the channel $(n_1,n_2)$ must be
\begin{equation}
 \langle\sigma_{(n_1,n_2)}\rangle=n_1\langle\sigma_1\rangle + n_2\langle\sigma_2\rangle + \delta_{(n_1,n_2)},
 \label{Model1Equation}
\end{equation}
where $\langle\hat{\sigma}_1\rangle=\sin(2\alpha)$,  $\langle\hat{\sigma}_2\rangle=\sin(2\beta)$, and $\alpha$ and $\beta$ are the angles between the fast axes of the waveplates and the initial linear driver polarizations. 

Here each of the three angular momenta can be measured independently, both experimentally and numerically, and thus a deviation term $\delta_{(n_1,n_2)}$ has been introduced for consistency. Within Model~1, the harmonic generation process is parametric if and only if this term is zero. Fleischer \etal{} attribute deviations from this to the failure of perturbative nonlinear optics and the presence of additional excitations, and call $\delta_{(n_1,n_2)}$ a `strong field correction'. Model~1 makes multiple predictions which agree with the experiment, though some of them require nonzero values of $\delta_{(n_1,n_2)}$.

\begin{enumerate}

 \item 
 For the symmetric case that $\alpha=\beta=45\degree$, so $\sigma_1=1$ and $\sigma_2=-1$, setting $\delta_{(n_1,n_2)}=0$ turns the basic relation \eqref{Model1Equation} into $\sigma_{(n_1,n_2)}=n_1-n_2$. From here, imposing the boundedness of photon spins, 
 \begin{equation}
 |\sigma_{(n_1,n_2)}|\leq 1,
 \label{PhotonSpinBoundedness}
 \end{equation} coupled with the parity constraint, means that $n_1$ and $n_2$ must differ by unity, which matches the experimental predictions.
 
 \item
 As the fundamental driver's waveplate is rotated away from the symmetric case, this restriction must be expanded to include the magnitude of $\sigma_2$, and now reads
 \begin{equation}
 |n_1\sin(2\alpha) - n_2| \leq 1.
 \label{ChannelExistenceRegion}
 \end{equation}
 For each channel $n_1$ and $n_2$ are fixed, so this reads as a restriction on $\alpha$, and gives the region where the channel is allowed:
 \begin{equation}
 \frac12\arcsin\left(\frac{n_2-1}{n_1}\right)\leq\alpha\leq\frac12\arcsin\left(\frac{n_2+1}{n_1}\right).
 \label{ChannelExistenceRegionUnbundled}
 \end{equation}
 This region matches well the observed range of certain channels, such as (7,6), (8,7), and (9,8). For certain series of channels, like (13,4), (12,5), (11,6), (10,7) and (9,8), this restriction also correctly predicts a V-shaped pattern where decreasing harmonic order gives an allowed region further from $\alpha=45\degree$. On the other hand, to obtain the correct regions, correction factors as high as $|\delta_{(n_1,n_2)}|=3$ are required, and these are not consistent across these channels \cite[see][supplementary information]{FleischerCohen}.
 
 \item \label{DisallowedChannelsDiscontinuity}
 For certain channels like (6,7) or (7,8), setting $\delta_{(n_1,n_2)}$ to zero makes the restriction \eqref{ChannelExistenceRegion} take the form
 \begin{equation}
 \sin(2\alpha) \geq 1.
 \label{ChannelExistenceDiscontinuity}
 \end{equation}
 This implies that parametric channels of this form are only allowed for $\alpha=45\degree$, but not for any nearby angles. This discontinuity is not present elsewhere in the formalism, and it is not observed in experiment or in simulations, so one is forced, within Model~1, to abandon conservation of spin angular momentum in the generation each individual harmonic.
 
 \item
 In its form $\frac{n_2-1}{n_1}\leq\sin(2\alpha)$, the restriction \eqref{ChannelExistenceRegion} means that, for $\beta$ fixed at 45\degree, only channels with $n_1\geq n_2-1$ can exist, which is in agreement with experiment.

\end{enumerate}

\newlength{\figheight}
\setlength{\figheight}{0.392 \textheight}
\newlength{\figwidth}
\setlength{\figwidth}{0.8\figheight}
\begin{figure*}
\centering
\subfigure[\ Right-circular harmonics]{ %
  \includegraphics[height=\figheight]{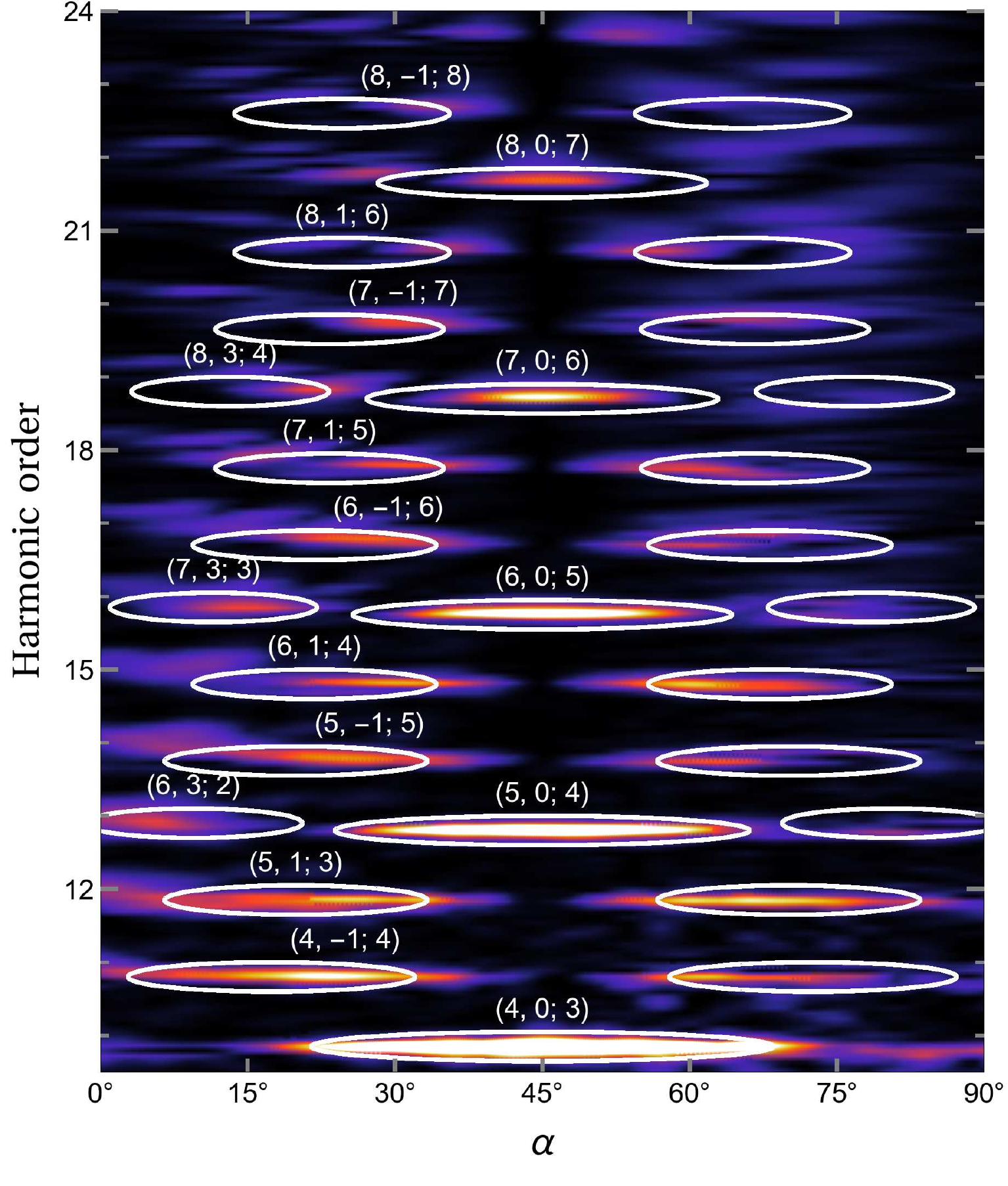} %
} %
\subfigure[\ Left-circular harmonics]{ %
  \includegraphics[height=\figheight]{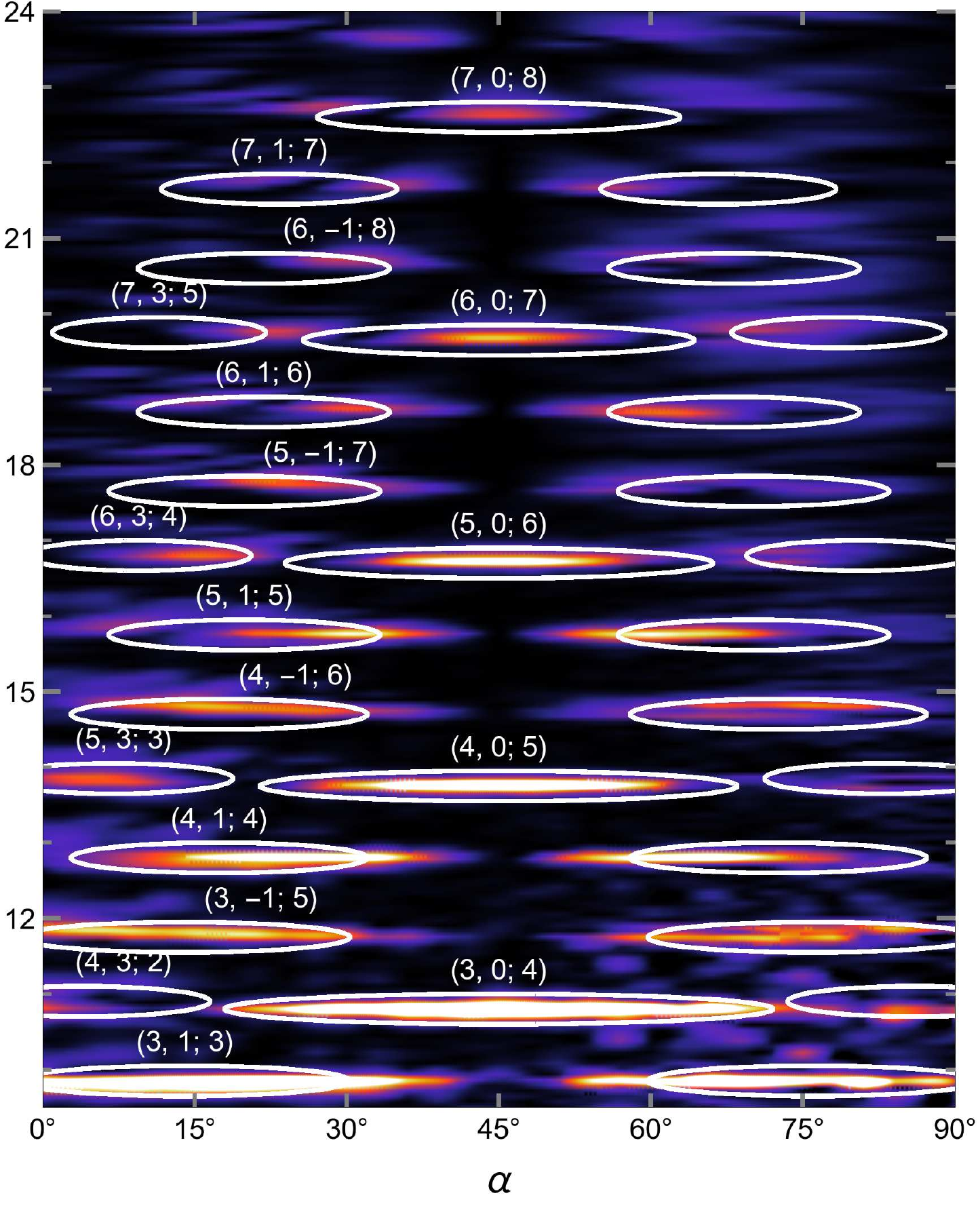} %
}%
\subfigure{
  \includegraphics[height=\figheight]{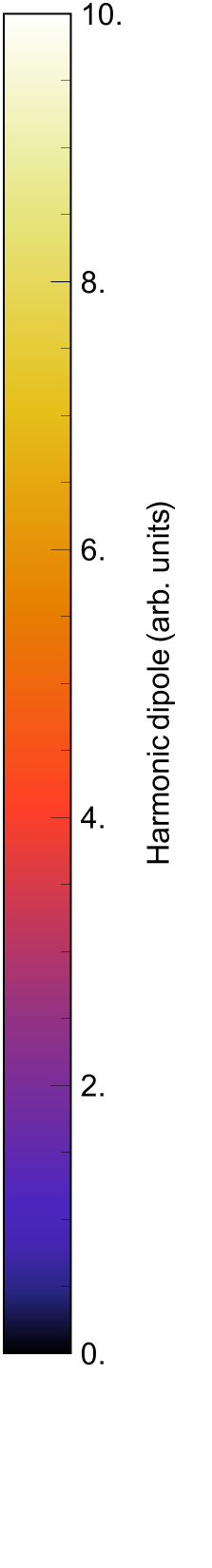} %
}
\caption{
  Existence regions for the different harmonics predicted by Model~2 compared to numerical simulations. The ellipses are drawn with arbitrary widths at the half-maximum-intensity ranges in ellipticity defined by Eq.~\eqref{BasicIntensityAlphaDependence}. We display only the lowest-order channel for each harmonic order and helicity, though higher-order channels are also present which partly overlap with the ones displayed. The background spectra are two-dimensional numerical simulation results for a 40-fs sine-squared pulse, with no detuning in the second harmonic, using a model potential for argon at equal driver intensities of $2\times10^{14}\,\text{W}/\text{cm}^2$.
}
\label{fig:ExistenceRegionEllipses}
\end{figure*}

Finally, within this model it is possible to study the deviation $\delta_{(n_1,n_2)}$ as a function of the experimental parameters. It is shown in Ref.~\citealp{FleischerCohen} that the average of this quantity over all the channels tends to be close to zero, which would indicate the possibility that harmonics are emitted in pairs, with the production of each pair conserving angular momentum. This is indeed possible, in principle, and in such a process Eq.~\eqref{Model1Equation} would be replaced by a more general conservation law for the two correlated channels seen as a single process. However, this picture does require a re-understanding of the three-step model.

\section{Decomposition-based model}
\label{sec:model}
We now present an alternative model for this experiment, which explains the above features while still allowing for the generation of each harmonic to preserve spin angular momentum independently of the other channels. The key to this model is seeing Eq.~\eqref{EllipticalAsSumOfCircularPolarizations} as indicating the presence of a third wave which must be included as such, instead of a change to the angular momentum carried by each photon of the driver.

To bring this to the forefront, we rephrase Eq.~\eqref{EllipticalAsSumOfCircularPolarizations} in the form
\begin{equation}
 \vbe=
 \frac{E_0e^{-i\omega t}}{2}
 \left(
 \cos(\delta\alpha)\une_R
 +
  \sin(\delta\alpha)\une_L
 \right)
 +\cc,
 \label{EllipticalAsTwoFields}
\end{equation}
where $\delta\alpha=\alpha-\pi/4$ and we have used $\eps=\tan(\alpha)$. We focus for simplicity on the case where $\beta$ is fixed at 45\degree.

Within Model~2, the problem consists now of \textit{three} waves which can combine to form harmonics: a left-circular harmonic driver at frequency $r\omega=1.95\omega,$ and two fundamental drivers at frequency $\omega$, one right-circular with relative amplitude $\cos(\delta\alpha)$, and one left-circular with relative amplitude $\sin(\delta\alpha)$. Each channel is now characterized by three integers, $(n_+,n_-;n_2)$, where $n_+$ ($n_-$) photons are absorbed from the right- (left-)circular fundamental driver, and $n_2$ from the harmonic driver, to give an emitted frequency of
\begin{equation}
 \Omega_{(n_+,n_-;n_2)}=(n_+ + n_-)\omega+n_2r\omega.
 \label{Model2EnergyConservation}
\end{equation}

Certain channels require negative values for $n_-$ or $n_+$ for one or both spins of the harmonic photon. In this case, the channel represents stimulated emission into that driver. This is necessary, for example, to explain the observed generation of elliptically polarized photons on channels of the form $(n_1,n_1+1)$ like $(6,7)$ and $(7,8)$. This is, however, not particularly surprising, because in this extreme nonlinear setting each harmonic contains contributions from processes of very many orders, and all but the lowest of these contain absorption and stimulated re-emission of photons from and to the driver fields.

Since each field has photons of a definite spin, the conservation of angular momentum reads in this model as
\begin{equation}
 \sigma_{(n_+,n_-;n_2)}=n_+\sigma_+ + n_-\sigma_- + n_2\sigma_2,
 \label{Model2AngularMomentumConservation}
\end{equation}
where $\sigma_+=+1$ and $\sigma_-=\sigma_2=-1$.

To obtain predictions, we apply the basic principle that the amplitude of an $n$-photon process should scale as the $n^\text{th}$ power of the driving field. This describes the leading term in the corresponding perturbation expansion, and applies both to absorption and to stimulated emission.

As the waveplate is rotated away from the symmetric setting at $\alpha=45\degree$, the initial energy is transferred from the right-circular driver to the left-circular one. Each channel $(n_+,n_-;n_2)$ absorbs an independent number of photons from each driver, which means that its amplitude must have a basic dependence of the form
\begin{equation}
 E_{(n_+,n_-,n_2)}\sim\cos^{|n_+|}(\delta\alpha)\sin^{|n_-|}(\delta\alpha),
 \label{BasicAlphaDependence}
\end{equation}
and the harmonic intensity is the square of this,
\begin{equation}
 I_{(n_+,n_-;n_2)}\sim\cos^{2|n_+|}(\delta\alpha)\sin^{2|n_-|}(\delta\alpha).
 \label{BasicIntensityAlphaDependence}
\end{equation}
For most channels $n_+$ and $n_-$ are relatively large integers, so the functions in Eqs.~\eqref{BasicAlphaDependence} and \eqref{BasicIntensityAlphaDependence} can be rather sharply peaked. 

Within this model there are no hard boundaries to the existence regions, and the harmonics are in principle possible for any set of laser parameters. Instead, the predictions are in terms for the basic profile of each channel as a function of the driver ellipticity.

A good approximation to where each channel is relevant is the region where it is above half of its maximum intensity; we display these regions in Fig.~\ref{fig:ExistenceRegionEllipses}. One interesting feature of this model is that each channel splits into two different channels with opposite spin. For instance, the channel identified as $(10, 5)$ in Model~1, at frequency $\Omega=(10+5r)\omega$, splits into the two channels $(8, 2; 5)$ and $(7, 3; 5)$, with spin $+1$ and $-1$ respectively. In general, the channel $(n_1,n_2)$ splits into the channels
\begin{equation}
 (n_+,n_-,n_2)=\left(\frac{n_1+n_2+\sigma}{2},\frac{n_1-n_2-\sigma}{2};n_2\right)
 \label{ChannelTranslation}
\end{equation}
with spin $\sigma=\pm1$. For this expression to give integer $n_\pm$, $n_1+n_2$ must be an odd integer, which matches the parity constraint of Model~1.

As is seen in Fig.~\ref{fig:ExistenceRegionEllipses}, the existence regions for these two channels overlap but do not coincide, and they agree rather well with numerical simulations without any free parameters. The superposition of right- and left-circular contributions whose amplitude peaks at different driver ellipticities helps explain the rich dynamics of the polarization of each harmonic shown by both experiment and numerics. 

One particularly important feature of this model is its behaviour for channels of the form $(n_1,0;n_1+1)$, like $(6,0;7)$. As remarked in point \ref{DisallowedChannelsDiscontinuity} above, conservation of angular momentum closes this channel within Model~1 for $\alpha\neq45\degree$: the second-harmonic driver contributes $-7$ units of angular momentum, and the six spins of $\sin(2\alpha)$ are only sufficient to allow a physical harmonic spin of $\sigma>-1$ when $\sin(2\alpha)=1$. Within Model~2, on the other hand, a slightly off-circular field can still produce harmonics: it is seen as a circular field of slightly reduced intensity, with the added presence of a left-circular driver which cannot participate in the process at that order, so the harmonic signal is only reduced slightly.

The other predictions of Model~1 can also be replicated. The symmetric case is identical for both models, so the restriction that $|n_1+n_2|=1$ there also holds; the V-shaped pattern is explained well together with the existence regions of the harmonics; and the restriction that $n_1\geq n_2-1$ is a consequence of the identity $n_+=n_- + n_2 + \sigma$.

It should be stressed, however, that modelling HHG with lowest-order perturbation theory has intrinsic limitations, such as the complete lack of a harmonic plateau. In this extreme nonlinear setting, many orders of perturbation theory contribute to each harmonic, involving many steps of absorption and stimulated emission of driver photons, and there is as yet no consistent theory to account for their interference. Nevertheless, the basic ellipticity dependence of the lowest order, embodied in Eqs.~\eqref{BasicAlphaDependence} and \eqref{BasicIntensityAlphaDependence}, is a good guide to where to look for each channel; as we have seen, it is remarkably successful.

\section{Subchannel splittings}
\label{sec:SubchannelSplittings}
We see, then, that Model~2 can account well for the main features seen in the experiment and in numerical simulations. However, because of its limitations, it is desirable to have additional confirmation that it is indeed the correct way to understand the process.

One way to do this is to exploit the principle that the right- and left-circular components of an elliptical field must be treated independently by actually tuning their frequencies independently. That is, to modify the field in Eq.~\eqref{EllipticalAsTwoFields} into the form
\begin{equation}
 \vbe=
 \frac{E_0}{2}
 \left(
 \cos(\delta\alpha)e^{-i\omega t}\une_R
 +
  \sin(\delta\alpha)e^{-i\omega' t}\une_L
 \right)
 +\cc,
 \label{DetunedEllipticalField}
\end{equation}
where the frequency $\omega'$ of the counter-rotating fundamental is now independent of $\omega$. In such a field, the energy conservation equation reads
\begin{equation}
 \Omega_{(n_+,n_-;n_2)}=n_+\omega + n_-\omega'+n_2r\omega,
 \label{ModifiedEnergyConservation}
\end{equation}
and the old channels $(n_1,n_2)$ should split into the two subchannels of Eq.~\eqref{ChannelTranslation} with a splitting proportional to the detuning $\delta=\omega'-\omega$.

In the time domain, the field in \eqref{DetunedEllipticalField} has an elliptical polarization which slowly rotates over time, since the two circular components, at close to the same frequency, accumulate a relative phase throughout the pulse. The axes of this ellipse must perform at least one full rotation: for a splitting of $\delta$ to be detected in the spectrum, the harmonic linewidth must be of that order, which means the pulse must be longer than $2\pi/\delta$, and therefore the two circular components accumulate a relative phase of at least $2\pi$ over the whole pulse.

This variation on the experiment can, in principle, be tested experimentally, though this adds a further layer of complexity. On the other hand, it is straightforward to implement numerically and it does not add new complications to the numerical methods, which must already be general enough to deal with arbitrary polarizations in two dimensions.

\setlength{\figheight}{0.395 \textheight}
\setlength{\figwidth}{0.8\figheight}
\begin{figure*}
\centering
\subfigure[\ Right-circular harmonics]{ %
  \includegraphics[height=\figheight]{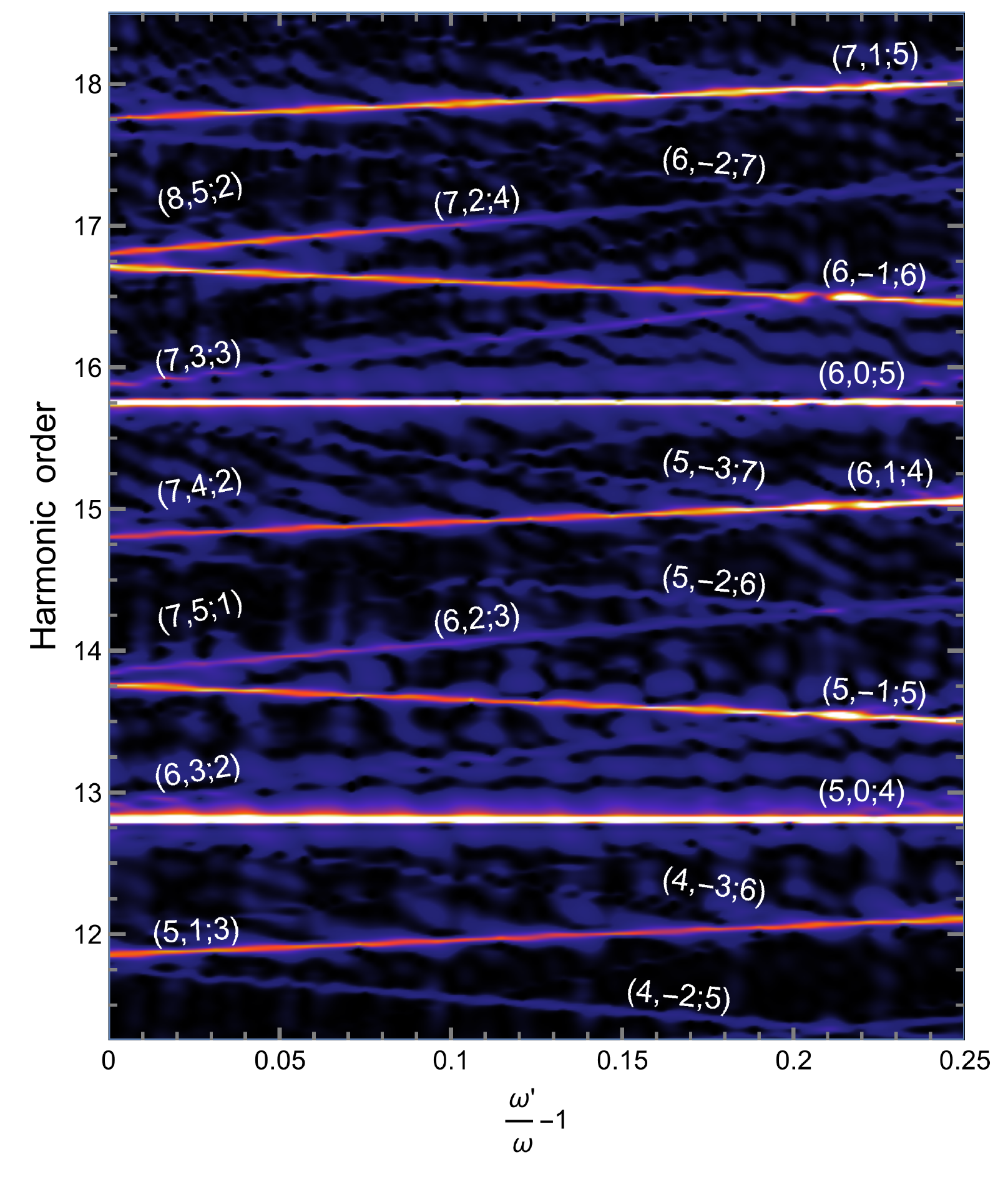} %
} %
\subfigure[\ Left-circular harmonics]{ %
  \includegraphics[height=\figheight]{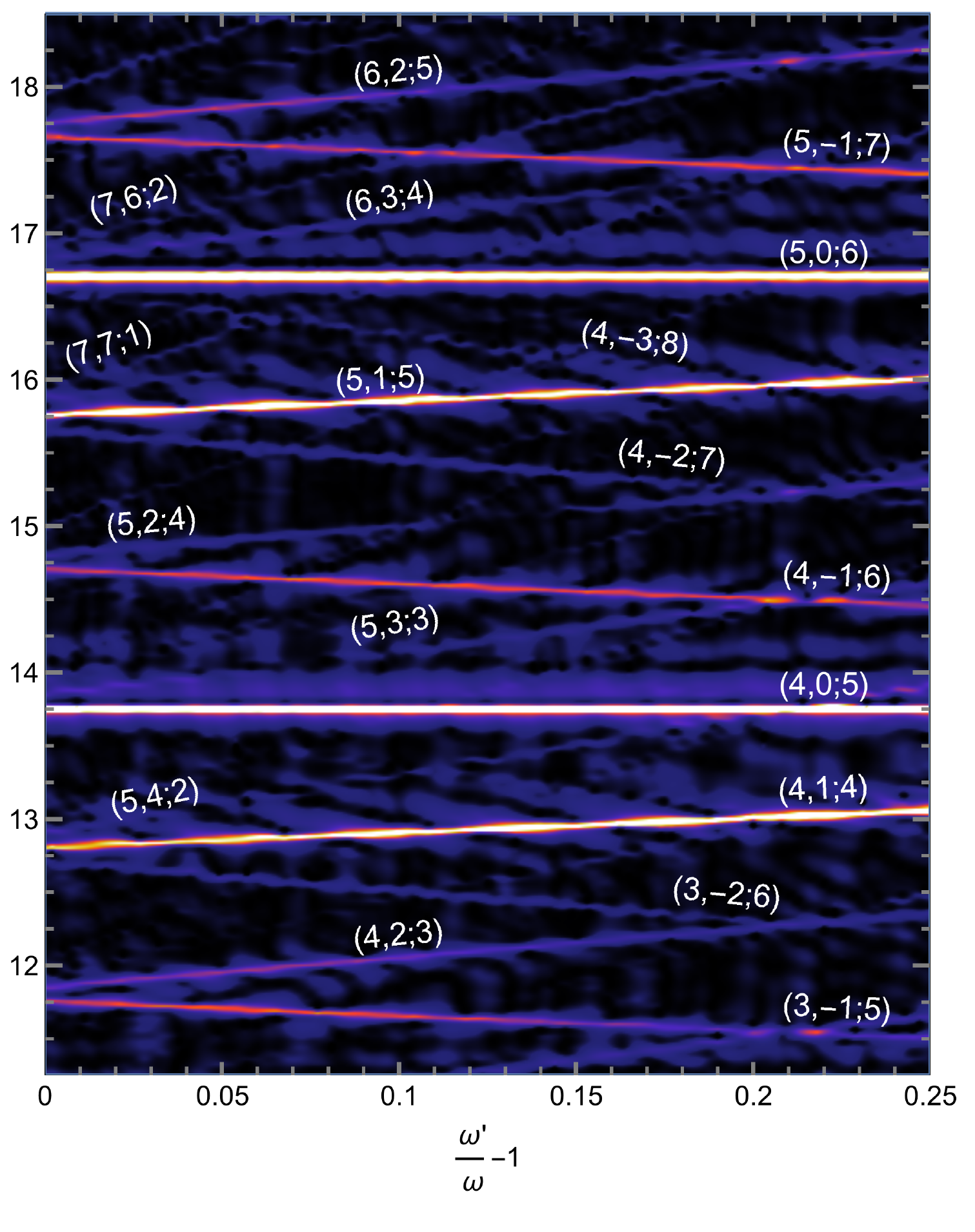} %
}%
\subfigure{
  \includegraphics[height=\figheight]{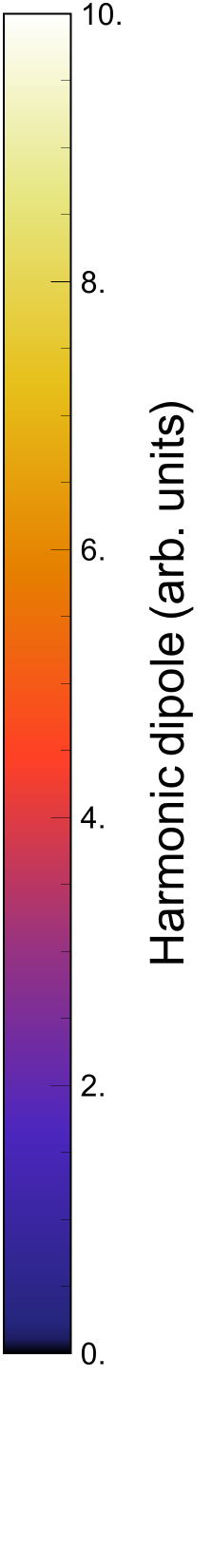} %
}
\caption{
  Dependence of the harmonic energies as a function of the relative detuning $\delta/\omega = \omega'/ \omega-1$ between the right- and left-circular components of the elliptically polarized fundamental driver is changed, as in Eq.~\eqref{ModifiedEnergyConservation}. We numerically integrate the SFA dipole \cite{LewensteinHHG,SFAcodeInGitHub}, and we use a flat-top pulse of 20 cycles of the fundamental with 2$\tfrac 12$ cycles of sinusoidal on- and off-ramp. We take equal intensities of $2\times10^{14}\,\text{W}/\text{cm}^2$ in the drivers, with $\alpha=35\degree$, and an S-type ground state with an argon-like $I_p$ of $15.6\,\text{eV}$.
}
\label{fig:SFA-splittings-spectrum}
\end{figure*}

As a test of this variation, then, we calculate the resulting spectra within the Strong Field Approximation (SFA), by direct numerical integration \cite{LewensteinHHG,SFAcodeInGitHub}. These results are shown in Fig.~\ref{fig:SFA-splittings-spectrum}, and they show the correct linear dependence of the harmonic energy $\Omega_{(n_+,n_-;n_2)}$ as a function of the relative detuning $\delta/\omega$ between the two circular components of the fundamental. Subchannels with as many as seven photons absorbed from the left-circular component can be identified, even though, at $\alpha=35\degree$, the counter-rotating component of the fundamental carries 3\% of the total intensity.

While it is clear that there are additional mechanisms and higher-order channels at work (as shown, particularly, by the intensity modulations of the harmonic lines over detuning), the harmonic energies follow very tightly the essential linear dependence with the correct slopes. This is strong evidence that the photon-exchange picture of Model~2 is the correct way of interpreting the experiment, both in the detuned cases and in the degenerate case of pure elliptical polarization, when $\omega'=\omega$.

\section{The four-wave mixing case}
\label{sec:perturbativesetting}
Having reviewed both models, we now focus on the lowest-order channel, which reduces to sum-frequency generation in standard four-wave mixing. This process is possible at much lower intensities, where ionized electrons cannot carry away angular momentum, so this brings the problems of Model~1 to the fore. This also means that the standard methods of perturbative nonlinear optics are applicable, and we show that this coincides with the predictions of Model~2.

\begin{figure}[ht]
 \centerline{
   \begin{tikzpicture}[
      scale=0.5,
      level/.style={thick},
      photon/.style={thick,->,shorten >=0.5pt,shorten <=0.5pt,>=stealth},
    ]    
    \newlength{\smap} \setlength{\smap}{2cm} 
    \newlength{\bigp} \setlength{\bigp}{3cm} 
    \newlength{\offset} \setlength{\offset}{6cm} 
    \newlength{\sep} \setlength{\sep}{0.35cm} 
    \newlength{\lvlwidth} \setlength{\lvlwidth}{0.75cm} 
    \newlength{\cancelm} \setlength{\cancelm}{0.82cm} 
    \newlength{\canceld} \setlength{\canceld}{0.75cm} 
    \newlength{\cancelv} \setlength{\cancelv}{1.7cm} 
    \draw[level] (-\offset-\lvlwidth,  0cm) -- (-\offset+\lvlwidth,  0cm);
    \draw[level] (-\offset-\lvlwidth,\smap+2\bigp) -- (-\offset+\lvlwidth,\smap+2\bigp);
    \draw[photon] (-\offset-\sep,    0cm) -- (-\offset-\sep, \smap) node[midway, left] {\begin{turn}{90}$\omega_1,\,\eps$\epol\end{turn}};
    \draw[photon] (-\offset-\sep,  \smap) -- (-\offset-\sep,\smap+\bigp) node[midway, left] {\begin{turn}{90}$\omega_2,\leftpol$\end{turn}};
    \draw[photon] (-\offset-\sep, \smap+\bigp) -- (-\offset-\sep,\smap+2\bigp) node[midway, left] {
       \begin{turn}{90}$\omega_2,\leftpol$ \end{turn}
       };
    \draw[photon] (-\offset+\sep,\smap+2\bigp) -- (-\offset+\sep,  0cm) node[midway,right] {\begin{turn}{90}$\omega_1+ 2\omega_2,$ ?\end{turn}};
    \node at (-\offset/2-1.0cm,\smap/2+\bigp-0.09cm) {\scalebox{1.5}{$=$}};
    \node at (-\offset/2+0.5cm,\smap/2+\bigp) {$\cos(\delta\alpha)$};
    \draw[level] (-\lvlwidth,  0cm) -- (\lvlwidth,  0cm);
    \draw[level] (-\lvlwidth,\smap+2\bigp) -- (\lvlwidth,\smap+2\bigp);
    \draw[photon] (-\sep,    0cm) -- (-\sep, \smap) node[midway, left] {\begin{turn}{90}$\omega_1,\rightpol$\end{turn}};
    \draw[photon] (-\sep,  \smap) -- (-\sep,\smap+\bigp) node[midway, left] {\begin{turn}{90}$\omega_2,\leftpol$\end{turn}};
    \draw[photon] (-\sep, \smap+\bigp) -- (-\sep,\smap+2\bigp) node[midway, left] {\begin{turn}{90}$\omega_2,\leftpol$ \end{turn}};
    \draw[photon] (\sep,\smap+2\bigp) -- (\sep,  0cm) node[midway,right] {\begin{turn}{90}$\omega_1+ 2\omega_2,\leftpol$\end{turn}};
    \node at (\offset/2-1.0cm,\smap/2+\bigp-0.006cm) {\scalebox{1.5}{$+$}};
    \node at (\offset/2+0.5cm,\smap/2+\bigp) {$\sin(\delta\alpha)$};
    \draw[level] (\offset-\lvlwidth,  0cm) -- (\offset+\lvlwidth,  0cm);
    \draw[level] (\offset-\lvlwidth,\smap+2\bigp) -- (\offset+\lvlwidth,\smap+2\bigp);
    \draw[photon] (\offset-\sep,    0cm) -- (\offset-\sep, \smap) node[midway, left] {\begin{turn}{90}$\omega_1,\leftpol$\end{turn}};
    \draw[photon] (\offset-\sep,  \smap) -- (\offset-\sep,\smap+\bigp) node[midway, left] {\begin{turn}{90}$\omega_2,\leftpol$\end{turn}};
    \draw[photon] (\offset-\sep, \smap+\bigp) -- (\offset-\sep,\smap+2\bigp) node[midway, left] {\begin{turn}{90}$\omega_2,\leftpol$\end{turn}};
    \draw[photon] (\offset+\sep,\smap+2\bigp) -- (\offset+\sep,  0cm) node[midway,right] {\begin{turn}{90}$\omega_1+ 2\omega_2, \,3\!\leftpol$\end{turn}};
    \draw[thick] (\offset+\cancelm - \canceld, \smap/2+\bigp - \cancelv) -- (\offset+\cancelm + \canceld, \smap/2+\bigp + \cancelv);
    \draw[thick] (\offset+\cancelm + \canceld, \smap/2+\bigp - \cancelv) -- (\offset+\cancelm - \canceld, \smap/2+\bigp + \cancelv);
    \end{tikzpicture}
 }
 \caption{
   The channel (1,2), where an elliptical driver at frequency $\omega_1$ and a left-circular driver at frequency $\omega_2$ produce a left-circular harmonic at frequency $\omega_1+2\omega_2$, can be understood as a simple four-wave mixing process. It can thus be treated perturbatively, and it will occur at much lower intensities.By decomposing the elliptical driver as a superposition of circular polarizations, as in Eqs.~\eqref{EllipticalAsSumOfCircularPolarizations} and~\eqref{EllipticalAsTwoFields}, one obtains an allowed process with a right-circular $\omega_2$ driver, and a forbidden process with three left-circular drivers which has too much angular momentum for a single harmonic photon.
 }
 \label{fig:FourWaveMixingScheme}
\end{figure}
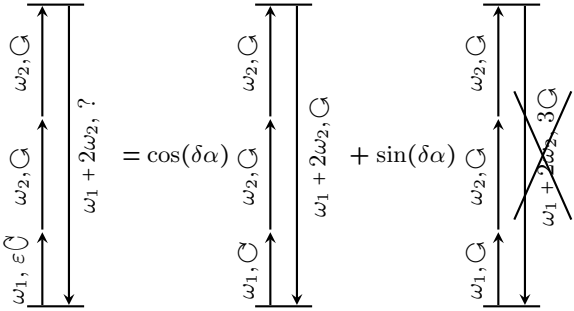

Consider, then, the channel (1,2), which is of the problematic form $(n_1,n_1+1)$ discussed in point \ref{DisallowedChannelsDiscontinuity} above. This is essentially the generation of the sum frequency $\omega_3=\omega_1+2\omega_2$ \cite{BloembergenSecondHarmonic}, and it can be done at much lower intensities in any medium with an isotropic third-order susceptibility tensor $\chit$; it is shown schematically in Fig.~\ref{fig:FourWaveMixingScheme}. As before, the driver at $\omega_2=r\omega$ is fixed at a left circular polarization, while the ellipticity $\eps$ of the driver at $\omega_1=\omega$ can be varied from right circular through linear to left circular.

From the perspective of Model~1, the process cannot happen unless the $\omega_1$ driver has a right circular polarization, with an ellipticity of $\eps=1$. If the field is even slightly elliptical, the expectation value of the spin per photon decreases to $\langle\sigma_1\rangle  =2\eps/(1+\eps^2) =\sin(2\alpha) <1$, and there is no longer a way for the total spin to be greater than -1.

Within Model~2, on the other hand, the elliptical driver is understood as a superposition of circular drivers of spin $\pm1$ with amplitude $(1\pm\eps)/\sqrt{2(1+\eps^2)}$. If the polarization is slightly off-circular, most of the amplitude is in the right-circular driver, which can still participate in the process, and a slightly reduced harmonic signal is obtained. 

More specifically, as the allowed process takes in one photon from the right-circular component at frequency $\omega_1$, the harmonic field will be proportional to the component's amplitude,
\begin{equation}
E\sim\frac{1+\eps}{\sqrt{2(1+\eps^2)}}=\cos(\delta\alpha),
\end{equation}
and the output intensity will be the square of this. Note, in particular, that there will be some nonzero harmonic intensity for \textit{all} ellipticities except for the completely left-circular case, which includes many cases with negative expectation value of the photon spin.

The predictions of Model~2 are in complete agreement with the predictions of standard perturbative nonlinear optics \cite{VectorFourWaveMixing, AgrawalFiberOptics}, which was shown early on to conserve spin angular momentum \cite{BloembergenConservationLaws,SelectrionRulesNonlinearOptics}. In this treatment, the sum-frequency wave at $\omega_3=\omega_1+2\omega_2$ is generated by the nonlinear polarization 
\begin{equation}
\vbp^{(3)}=\epsilon_0\chit
\tdots
\vbe\,\vbe\,\vbe,
\label{NonLinearPolarizationDefinition}
\end{equation}
where the vertical dots denote a three-way tensor contraction. In component form, this relation reads $P^{(3)}_i=\epsilon_0 \sum_{jkl} \chi^{(3)}_{ijkl} E_jE_kE_l$.

To obtain the sum-frequency component of this polarization, one expresses the electric field as a sum over the participating modes, 
\begin{equation}
  \vbe=\sum_{\alpha=1}^3\left[
  \vbe_\alpha e^{i(\vbk_\alpha\cdot \vbr-\omega_\alpha t)}
  +
  \vbe_\alpha^\ast e^{-i(\vbk_\alpha\cdot \vbr-\omega_\alpha t)}
  \right]
  \label{ElectricFieldWithConjugate}
\end{equation}
and looks for the component of the polarization which oscillates as $e^{i(\vbk_3\cdot\vbr-\omega_3 t)}$. Substituting the expression~\eqref{ElectricFieldWithConjugate} into the contraction in \eqref{NonLinearPolarizationDefinition} results in eight terms, depending on whether $\vbe_\alpha$ or its conjugate is taken. Each of the eight terms describes a different process, which include parametric amplification or self- and cross-phase modulation \cite{AgrawalFiberOptics}; the sum-frequency generation process we want is the term with three factors of  $\vbe_\alpha$. This has the polarization amplitude
\begin{equation}
\vbp_3=\epsilon_0\chit
\tdots
\vbe_1\vbe_2\vbe_2 e^{i\varphi},
\label{NonLinearPolarizationAmplitudes}
\end{equation}
where $\varphi=(\vbk_1+2\vbk_2-\vbk_3)\cdot\vbr-(\omega_1+2\omega_2-\omega_3)t$.

To calculate the contraction in Eq.~\eqref{NonLinearPolarizationAmplitudes} we impose the isotropy condition on the susceptibility tensor $\chit$. The only isotropic tensors of rank 4 have a component form $\delta_{ij}\delta_{kl}$ \cite[\S3.03]{JeffreysJeffreys}, which corresponds to the tensor action 
$$\stackrel{\leftrightarrow}{\delta}\, \,\tdots\vb{u}\vb{v}\vb{w}=\vb{u}(\vb{v}\cdot\vb{w}).$$
That is, the tensor contracts its second and third inputs, and produces a vector along its first input. The contraction in \eqref{NonLinearPolarizationAmplitudes} produces three terms of this form, with different permutations of its inputs. Each of these terms will in principle have a different frequency-dependent complex scalar susceptibility $\chi^{(3)}_\text{s}(\omega_\alpha,\omega_\beta,\omega_\gamma)$, but only one term will be allowed so this distinction can be dropped.

Under these conditions, the sum-frequency polarization becomes
\vspace{-1mm} 
\begin{equation}
\vbp_3=\epsilon_0\chis e^{i\varphi} \left(\vphantom{\sum_i}
2\vbe_2(\vbe_1\cdot\vbe_2)+\vbe_1(\vbe_2\cdot\vbe_2)
\right).
\label{NonLinearPolarizationInnerProducts}
\end{equation}
Here $\vbe_2=E_2\une_L$ is left polarized, which means that the second term vanishes: in a frame where the propagation direction is in the $z$ axis,
\begin{equation}
 \une_L\cdot\une_L=
 \frac{1}{\sqrt{2}}\begin{pmatrix}1\\i\\0\end{pmatrix}
 \cdot
 \frac{1}{\sqrt{2}}\begin{pmatrix}1\\i\\0\end{pmatrix}
 =0.
\end{equation}
The amplitude for the field at $\omega_1$ encodes the ellipticity dependence, through the analog of Eq.~\eqref{EllipticalAsSumOfCircularPolarizations},
\begin{equation}
 \vbe_1=
 E_1
 \left(
 \frac{1+\eps}{\sqrt{2(1+\eps^2)}}\une_R
 +
  \frac{1-\eps}{\sqrt{2(1+\eps^2)}}\une_L
 \right).
 \label{EllipticalDriverPolarizationDecomposition}
\end{equation}
This is projected on the amplitude $\vbe_2$, and multiplies the left-circular vector $\vbe_2$, so that the final amplitude is
\begin{equation}
\vbp_3=\epsilon_0\chis e^{i\varphi} E_1E_2^2 \frac{1+\eps}{\sqrt{2(1+\eps^2)}}\une_L.
\label{NonLinearPolarizationFinalResult}
\end{equation}

The ellipticity dependence of this result is exactly that predicted by Model~2, whereas Model~1 predicts the process is forbidden except for $\eps=1$. Therefore, at least in the cases where perturbative optics holds, using the expectation value of each photon's angular momentum in the conservation equation leads to incorrect results. 

This is slightly counter-intuitive, as one does expect a conservation equation to hold at the level of expectation values for every conserved quantity, but a direct application in the form of Eq.~\eqref{Model1Equation} is inconsistent with formal perturbative treatments where those are available, and would need further justification for its use in more highly nonlinear cases.

Nevertheless, it is indeed possible to understand the generation of harmonics by elliptical drivers, in both the perturbative and extreme-nonlinear cases, in terms of a simple photon picture. Our model provides a simple framework for this understanding, which is in agreement with the available experimental observations and whose predictions are borne out by numerical calculations. The experiment of Fleischer~\etal~\cite{FleischerCohen}, then, is seen to be consistent with the conservation of spin angular momentum, and with a picture of HHG as a parametric process where multiple driver photons get up-converted into harmonic photons and the atom returns to its ground state after the recombination step.

We gratefully acknowledge fruitful discussions with A. Fleischer, O.~Cohen and O.~Kfir, who made their interpretations and results available prior to publication. This work was funded by EPSRC Program Grant\textsc{EP/I032517/\-1} and the CORINF Marie Curie Initial Training Network. E. P. thanks CO\-NA\-CYT for financial support. M.I. acknowledges partial support by the United States Air Force Office of Scientific Research under program No.~FA9550-12-1-0482.

\bibliography{references}{}

\begin{thebibliography}{10}
\providecommand{\url}[1]{\texttt{#1}}
\providecommand{\urlprefix}{URL }
\expandafter\ifx\csname urlstyle\endcsname\relax
  \providecommand{\doi}[1]{doi:\discretionary{}{}{}#1}\else
  \providecommand{\doi}{doi:\discretionary{}{}{}\begingroup
  \urlstyle{rm}\Url}\fi
\providecommand{\selectlanguage}[1]{\relax}
\providecommand{\bibAnnoteFile}[1]{%
  \IfFileExists{#1}{\begin{quotation}\noindent\textsc{Key:} #1\\
  \textsc{Annotation:}\ \input{#1}\end{quotation}}{}}
\providecommand{\bibAnnote}[2]{%
  \begin{quotation}\noindent\textsc{Key:} #1\\
  \textsc{Annotation:}\ #2\end{quotation}}
\providecommand{\eprint}[2][]{\url{#2}}

\bibitem{HHGTutorial}
\textsc{M.~Ivanov and O.~Smirnova}.
\newblock {M}ultielectron {H}igh {H}armonic {G}eneration: simple man on a
  complex plane{.
\newblock In \textsc{T.~Schultz and M.~Vrakking} (eds.), \emph{Attosecond and
  {F}ree {E}lectron {L}aser science}, chap.~1, pp. 1--55 (2013)}.
\newblock \href{http://arxiv.org/abs/1304.2413}{arXiv:1304.2413
 } [physics.atom-ph].
\bibAnnoteFile{HHGTutorial}
\bibitem{JoachainReview}
\textsc{C.~Joachain, M. D{\"o}rr and N.~Kylstra}.
\newblock High-intensity laser-atom
  physics. \href{http://dx.doi.org/10.1016/S1049-250X(08)60188-3}{
\newblock \emph{Adv. At., Mol., Opt. Phys.} \textbf{42} (2000), pp. 225--286}.
\newblock \href{http://www.ulb.ac.be/sciences/physath/publi/advan_10.ps.gz}{ULB
  e-print}.
\bibAnnoteFile{JoachainReview}

\bibitem{UltrahighHG}
\textsc{T.~Popmintchev, M.-C. Chen, D.~Popmintchev} et~al.
\newblock Bright coherent ultrahigh harmonics in the {keV} {X}-ray regime from
  mid-infrared femtosecond
  lasers. \href{http://dx.doi.org/10.1126/science.1218497}{
\newblock \emph{Science} \textbf{336} no. 6086 (2012), pp. 1287--1291}.
\newblock
  \href{http://jila.colorado.edu/kmgroup/sites/default/files/pdf/Science-2012-Popmintchev-1287-91.pdf}{JILA
  e-print}.
\bibAnnoteFile{UltrahighHG}

\bibitem{EnergyConservationExperiment}
\textsc{M.~D. Perry and J.~K. Crane}.
\newblock High-order harmonic emission from mixed
  fields. \href{http://dx.doi.org/10.1103/PhysRevA.48.R4051}{
\newblock \emph{Phys. Rev. A} \textbf{48} no.~6 (1993), pp. R4051--R4054}.
\bibAnnoteFile{EnergyConservationExperiment}

\bibitem{MomentumConservationExperiment}
\textsc{J.~B. Bertrand, H.~J. W{\"o}rner, H.-C. Bandulet, {\'E}.~Bisson,
  M.~Spanner, J.-C. Kieffer, D.~M. Villeneuve and P.~B. Corkum}.
\newblock Ultrahigh-order wave mixing in noncollinear high harmonic
  generation. \href{http://dx.doi.org/10.1103/PhysRevLett.106.023001}{
\newblock \emph{Phys. Rev. Lett.} \textbf{106} no.~2 (2011), p. 023\,001}.
\bibAnnoteFile{MomentumConservationExperiment}

\bibitem{OAMConservationExperiment}
\textsc{M.~Z{\"u}rch, C.~Kern, P.~Hansinger, A.~Dreischuh and C.~Spielmann}.
\newblock Strong-field physics with singular light
  beams. \href{http://dx.doi.org/10.1038/nphys2397}{
\newblock \emph{Nature Phys.} \textbf{8} no.~10 (2012), pp. 743--746}.
\bibAnnoteFile{OAMConservationExperiment}

\bibitem{FleischerCohen}
\textsc{A.~Fleischer, O.~Kfir, T.~Diskin, P.~Sidorenko and O.~Cohen}.
\newblock Spin angular momentum and tunable polarization in high-harmonic
  generation. \href{http://dx.doi.org/10.1038/nphoton.2014.108}{
\newblock \emph{Nature Photon.} \textbf{8} no.~7}.
\newblock \href{http://arxiv.org/abs/1310.1206}{arXiv:1310.1206
 } [physics.optics].
\bibAnnoteFile{FleischerCohen}

\bibitem{EichmannExperiment}
\textsc{H.~Eichmann, A.~Egbert, S.~Nolte, C.~Momma, B.~Wellegehausen,
  W.~Becker, S.~Long and J.~K. McIver}.
\newblock Polarization-dependent high-order two-color
  mixing. \href{http://dx.doi.org/10.1103/PhysRevA.51.R3414}{
\newblock \emph{Phys. Rev. A} \textbf{51} (1995), pp. R3414--R3417}.
\bibAnnoteFile{EichmannExperiment}

\bibitem{SFALong}
\textsc{S.~Long, W.~Becker and J.~K. McIver}.
\newblock Model calculations of polarization-dependent two-color high-harmonic
  generation. \href{http://dx.doi.org/10.1103/PhysRevA.52.2262}{
\newblock \emph{Phys. Rev. A} \textbf{52} (1995), pp. 2262--2278}.
\bibAnnoteFile{SFALong}

\bibitem{SFAMilosevic}
\textsc{D.~B. Milo\ifmmode \check{s}\else \v{s}\fi{}evi\ifmmode~\acute{c}\else
  \'{c}\fi{} and B.~Piraux}.
\newblock High-order harmonic generation in bichromatic elliptically
  polarized laser field. \href{http://dx.doi.org/10.1103/PhysRevA.54.1522}{
\newblock \emph{Phys. Rev. A} \textbf{54} no.~2 (1996), pp. 1522--1531}.
\bibAnnoteFile{SFAMilosevic}

\bibitem{SFAMilosevicBecker}
\textsc{D.~B. Milo\ifmmode \check{s}\else \v{s}\fi{}evi\ifmmode~\acute{c}\else
  \'{c}\fi{}, W.~Becker and R.~Kopold}.
\newblock Generation of circularly polarized high-order harmonics by two-color
  coplanar field mixing. \href{http://dx.doi.org/10.1103/PhysRevA.61.063403}{
\newblock \emph{Phys. Rev. A} \textbf{61} no.~6 (2000), p. 063\,403}.
\bibAnnoteFile{SFAMilosevicBecker}

\bibitem{VitaliSelectionRules}
\textsc{V.~Averbukh, O.~E. Alon and N.~Moiseyev}.
\newblock Stability and instability of dipole selection rules for atomic
  high-order-harmonic-generation spectra in two-beam
  setups. \href{http://dx.doi.org/10.1103/PhysRevA.65.063402}{
\newblock \emph{Phys. Rev. A} \textbf{65} (2002), p. 063\,402}.
\newblock
  \href{http://tx.technion.ac.il/~nimrodhp/publications/ps/2002/va_oa_nm2002.pdf}{Technion
  e-print}.
\bibAnnoteFile{VitaliSelectionRules}

\bibitem{SFACeccherini}
\textsc{F.~Ceccherini, D.~Bauer and F.~Cornolti}.
\newblock Harmonic generation by atoms in circularly polarized two-color laser
  fields with coplanar polarizations and commensurate
  frequencies. \href{http://dx.doi.org/10.1103/PhysRevA.68.053402}{
\newblock \emph{Phys. Rev. A} \textbf{68} no.~5 (2003), p. 053\,402}.
\newblock \href{http://arxiv.org/abs/physics/0211110}{arXiv: physics/0211110
 } [physics.atom-ph].
\bibAnnoteFile{SFACeccherini}

\bibitem{MilosevicIsolatedPulses}
\textsc{D.~B. Milo\ifmmode \check{s}\else \v{s}\fi{}evi\ifmmode~\acute{c}\else
  \'{c}\fi{} and W.~Becker}.
\newblock Attosecond pulse generation by bicircular fields: from pulse trains
  to single pulse. \href{http://dx.doi.org/10.1080/09500340410001731011}{
\newblock \emph{Journal of Modern Optics} \textbf{52} no. 2-3 (2005), pp.
  233--241}.
\bibAnnoteFile{MilosevicIsolatedPulses}

\bibitem{LewensteinHHG}
\textsc{M.~Lewenstein, P.~Balcou, M.~Y. Ivanov, {A. L'Huil\-lier} and P.~B.
  Corkum}.
\newblock Theory of high-harmonic generation by low-frequency laser
  fields. \href{http://dx.doi.org/10.1103/PhysRevA.49.2117}{
\newblock \emph{Phys. Rev. A} \textbf{49} (1994), pp. 2117--2132}.
\bibAnnoteFile{LewensteinHHG}

\bibitem{SFAcodeInGitHub}
\textsc{E.~Pisanty}.
\newblock {RB-SFA}: {R}otating {B}icircular {H}igh {H}armonic {G}eneration in
  the {S}trong {F}ield {A}pproximation.
\newblock \url{https://github.com/episanty/RB-SFA},
  \href{http://dx.doi.org/10.5281/zenodo.11224}{v1.0} (2014).
\bibAnnoteFile{SFAcodeInGitHub}

\bibitem{BloembergenSecondHarmonic}
\textsc{H.~J. Simon and N.~Bloembergen}.
\newblock Second-harmonic light generation in crystals with natural optical
  activity. \href{http://dx.doi.org/10.1103/PhysRev.171.1104}{
\newblock \emph{Phys. Rev.} \textbf{171} no.~3 (1968), pp. 1104--1114}.
\bibAnnoteFile{BloembergenSecondHarmonic}

\bibitem{VectorFourWaveMixing}
\textsc{Q.~Lin and G.~P. Agrawal}.
\newblock Vector theory of four-wave mixing: polarization effects in
  fiber-optic parametric
  amplifiers. \href{http://dx.doi.org/10.1364/JOSAB.21.001216}{
\newblock \emph{J. Opt. Soc. Am. B} \textbf{21} no.~6 (2004), pp. 1216--1224}.
\bibAnnoteFile{VectorFourWaveMixing}

\bibitem{AgrawalFiberOptics}
\textsc{G.~P. Agrawal}.
\newblock \emph{Nonlinear Fiber Optics} (Academic Press, Boston, 1989).
\bibAnnoteFile{AgrawalFiberOptics}

\bibitem{BloembergenConservationLaws}
\textsc{N.~Bloembergen}.
\newblock Conservation laws in nonlinear
  optics. \href{http://dx.doi.org/10.1364/JOSA.70.001429}{
\newblock \emph{J. Opt. Soc. Am.} \textbf{70} no.~12 (1980), p. 1429}.
\bibAnnoteFile{BloembergenConservationLaws}

\bibitem{SelectrionRulesNonlinearOptics}
\textsc{C.~L. Tang and H.~Rabin}.
\newblock Selection rules for circularly polarized waves in nonlinear
  optics. \href{http://dx.doi.org/10.1103/PhysRevB.3.4025}{
\newblock \emph{Phys. Rev. B} \textbf{3} no.~12 (1971), pp. 4025--4034}.
\bibAnnoteFile{SelectrionRulesNonlinearOptics}

\bibitem{JeffreysJeffreys}
\textsc{H.~Jeffreys and B.~S. Jeffreys}.
\newblock \emph{Methods of Mathematical Physics}.
\newblock ed. (Cambridge University Press, 1965), p.~87.
\newblock
  \href{http://www.archive.org/details/MethodsOfMathematicalPhysics3rd.ed.}{ark:/13960/t3hx2br67}.
\bibAnnoteFile{JeffreysJeffreys}

\end{thebibliography}


\end{document}